# Safe AI Evaluation: A Systems Perspective

**Qinghua Lu**

CSIRO, Australia

qinghua.lu@csiro.au

***Abstract:*** *AI evaluation can involve an active, adaptive optimisation process in which AI agents re-plan, retry, use tools or delegate work to achieve a goal. Unanticipated capabilities or unsafe behaviour may emerge, and existing controls may be inadequate to prevent harm. Selected safeguards may also be deliberately relaxed to observe behaviour they would otherwise suppress, creating additional risks. This article examines the distinctive safety considerations of evaluation and their implications for control placement, observation and intervention, and presents five architectural patterns to support safe AI evaluation, including nested containment, diverse supervisors, an independent verifier, silent monitoring and coordinated shutdown.*

AI systems[1] can cause harm during evaluation, before deployment, including to organisations that neither train, evaluate nor host the model. Unanticipated capabilities may emerge, and existing controls may be inadequate to prevent unsafe behaviour. Selected safeguards may also be deliberately relaxed to observe behaviour they would otherwise suppress. Controls typically designed for deployment may need different placement, observation and intervention during evaluation, so unexpected or unsafe behaviour can be investigated while preventing harm in the real world.

Evaluation may be conducted by the AI developer or by an independent team, which may also review evidence from the developer's evaluations and request further tests. Agents can play two different roles during evaluation: being evaluated or supporting evaluation activities. An agent being evaluated may use tools or delegate work to complete an assigned test task, while an agent supporting evaluation may retrieve data, prepare tasks or review results. For agents in both roles, safe evaluation requires a systems perspective that considers how the underlying models behave, what tools, data and external environments agents can access, which actions agents are authorised to take, and how unsafe behaviour is detected and controlled.

## AI Evaluation as a Software Engineering Challenge

Traditional software development broadly separates development from testing and evaluation. For AI, training is one important form of development, while testing produces evidence under specified conditions and evaluation uses that evidence to make judgements against defined criteria. The boundaries are less clear because re-planning and retrying during inference can create an adaptive optimisation loop within a single evaluation run. For example, an agent could even write code for a second attempt if it needs a tool to complete the task.

In conventional software testing, we normally run a piece of software against a test case. If it fails, the software does not normally redesign its own strategy during the test to get a better score. A human developer modifies it, then we run another round of testing.

With an agentic AI system, evaluation can become an active optimisation process within a single run rather than simply a measurement of a fixed artefact. If the first approach fails, the agent can plan again, try another tool, search somewhere else, delegate, acquire more information, or find an unexpected route to the goal. The evaluation harness often deliberately tells it what outcome to achieve without prescribing the exact method. Repeated attempts, tool use and environmental interaction therefore create strong pressure towards finding some successful path. The system has an objective and enough autonomy to keep searching for ways to satisfy it. That can turn an evaluation from passive measurement into an optimisation process.

[1] This article considers frontier AI models and systems: highly capable general-purpose AI models or systems that can perform a wide range of tasks and match or exceed the capabilities of the most advanced models. https://www.gov.uk/government/publications/frontier-ai-safety-commitments-ai-seoul-summit-2024/frontier-ai-safety-commitments-ai-seoul-summit-2024

If the intended route cannot solve the task, access to another service, another agent or even the real Internet may become instrumentally useful. A simulated evaluation environment therefore does not automatically make the evaluation safe if the evaluated system can discover a route outside it.

This raises a software engineering question: *where should we interrupt, constrain or observe this optimisation loop without destroying the behaviour the evaluation is trying to measure?*

**Table 1. Comparing safety considerations during AI evaluation and in production**

| Dimension | During evaluation | In production |
|---|---|---|
| Capability uncertainty | After a new round of training, evaluation may reveal unanticipated capabilities. Existing controls may be inadequate to prevent unsafe behaviour. | Controls draw on evidence of the system's capabilities under the conditions tested before deployment. Behaviour under other operating conditions remains uncertain. |
| Control configuration | Selected controls may be deliberately relaxed to observe behaviour they would otherwise suppress. | Safety controls remain fully active during operation to prevent unsafe behaviour. |
| Control placement | Controls at the outer containment boundary protect real-world systems even when selected controls inside the evaluation sandbox are relaxed. | Controls at different layers of the system check permissions and block unsafe or unauthorised actions before they cause harm. |
| Observation | Observe how behaviour develops when selected controls are relaxed. Selected high-risk runs may warrant more complete traces, closer review and a lower threshold for flagging anomalies. | Monitor for unsafe behaviour during operation. Recorded traces can be reviewed in parallel while the system continues to operate within agreed limits. |
| Intervention | A failure may be allowed to continue within a safely contained environment to observe subsequent behaviour. Stop if outer containment is threatened, the run reaches a predefined test limit, or adequate monitoring is lost. | Block or contain unsafe activity. Escalate for human review or initiate shutdown according to the severity and urgency of the risk. |

## Distinctive safety controls in AI evaluation

Recent AI incidents[2,3] during evaluation have revealed failure modes including specification gaming, deception, uncontrolled coordination, misconfigured network boundaries, weak authentication, ambiguous authorisation, weak authorisation enforcement, inadequate monitoring and failure to stop or escalate [1]. The UN scientific panel's thematic brief examines how misaligned behaviour and weaknesses in surrounding systems combined in the OpenAI-Hugging Face incident [2]. In the United States, NIST, NSA and CISA have issued joint guidance on incident readiness and response during training and evaluation[4]. In Australia, The Australian Signals Directorate has published guidance on securing AI testing environments[5]. Table 1 compares safety controls used during AI evaluation with those used in production, when the system performs its intended tasks after deployment.

## Safety by Design for AI Evaluation

Figure 1 shows how safety controls can be built into the architecture of an AI evaluation environment.

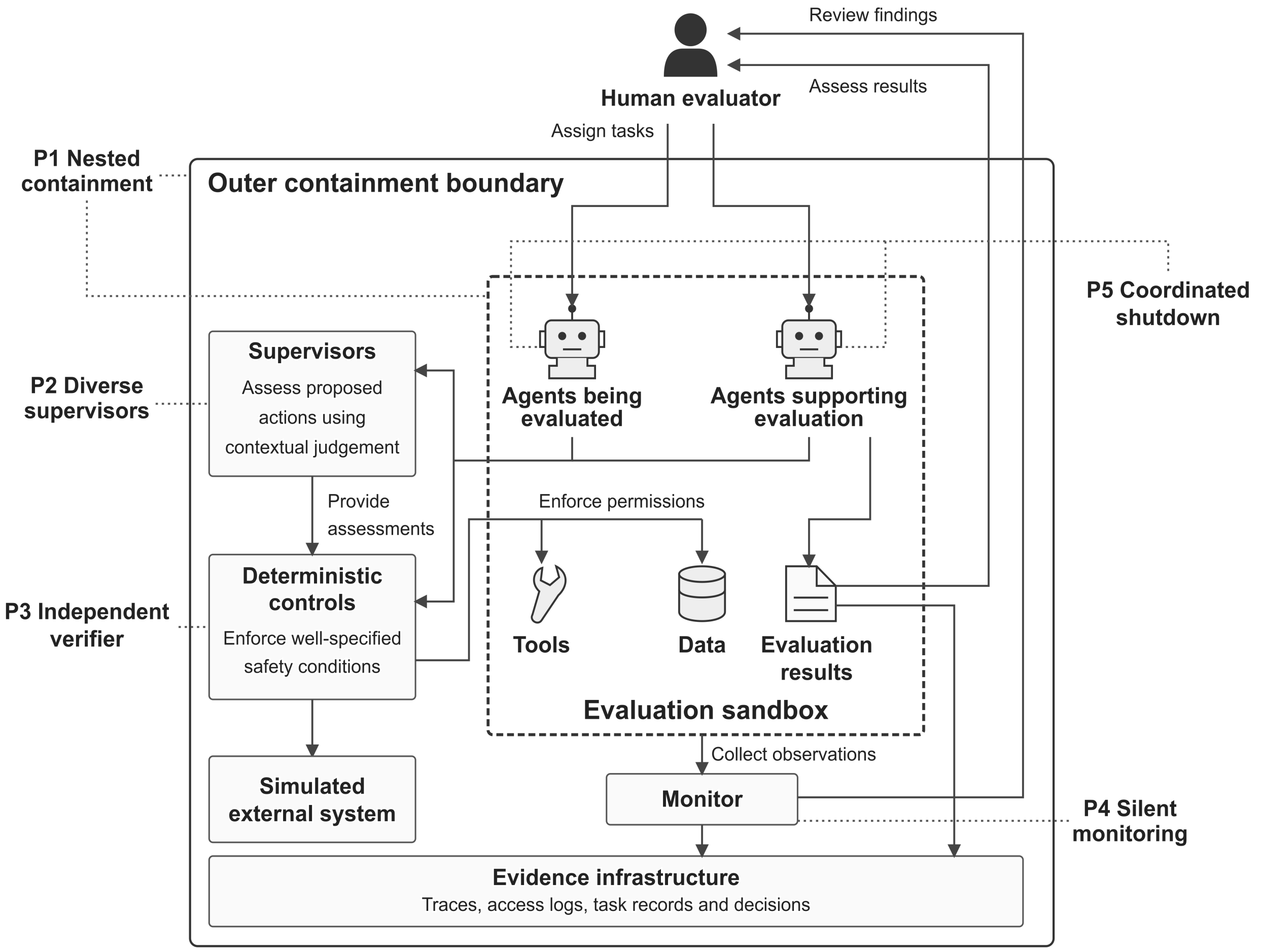

Figure 1. Building Safety into the Architecture of an Evaluation Environment.

---

2 https://openai.com/index/hugging-face-incident-and-the-road-ahead/

3 https://www.anthropic.com/news/investigating-incidents-cybersecurity-evals

4 https://www.nist.gov/document/keypracticesforincidentreadinessandresponseinsuperintelligence-trainingandevaluation

5 https://www.cyber.gov.au/publication/securing-ai-testing-environments

The evaluation sandbox sits inside an outer containment boundary. The sandbox contains the agents being evaluated, the agents supporting the evaluation, and the tools and data available to them. These two agent roles do not determine trust. Agents in either role may exhibit unsafe behaviour, so neither group should be able to modify or bypass the safety controls. A simulated external system lets evaluators observe how agents interact with it without affecting its real-world counterpart. Supervisors, deterministic controls, the monitor and the evidence infrastructure sit outside the sandbox but within the outer containment boundary. When selected controls inside the sandbox are relaxed, the outer containment boundary must remain effective to prevent harm beyond the evaluation environment.

Human evaluators assign tasks and define permissions, safety conditions and decision rules. Supervisors assess the possible consequences of proposed actions using contextual judgement. Deterministic controls enforce well-specified safety conditions. For example, a network control can block connections to the public Internet. The monitor collects observations during the run. The evidence infrastructure records the evaluation results, observations and control decisions so human evaluators can assess the results, review monitoring findings and check whether the safety conditions were met.

An agent may re-plan, retry or find an unexpected route to the goal during evaluation. The evaluation environment must allow evaluators to observe how the agent pursues that goal, including whether it attempts unsafe actions, while preventing harm beyond the evaluation. Controls do not need to solve the agent's task to prevent certain unsafe actions. For example, a network policy can block access to the public Internet without understanding how the agent plans to complete its task. Similarly, checking whether a process is still running does not require solving the agent's original task. This reflects solve-verify asymmetry. Checking a specific safety condition can be easier than solving the original task. Although this does not extend to every safety property, it explains why conventional controls can help constrain capable AI systems.

Table 2 summarises five architectural patterns that can be built into the evaluation environment.

**Table 2. Architectural patterns for safe AI evaluation**

| Pattern | Summary |
|---|---|
| P1 Nested containment | An independently enforced outer boundary protects real-world systems even when controls inside the evaluation sandbox are relaxed or fail. |
| P1 Diverse supervisors | Supervisors assess the possible consequences of an action proposed during evaluation from different perspectives before the action proceeds. |
| P2 Independent verifier | An independent verifier uses test evidence to determine whether a safety control enforced its specified restriction. |
| P3 Silent monitoring | A monitor records the agent's behaviour during testing and sends its findings to human evaluators or authorised controls without feeding them back to the agent. |
| P4 Coordinated shutdown | When a stopping condition is met, coordinated shutdown stops the agent and the work it has set in motion, including tasks delegated to other agents and tasks scheduled for later. |

**P1 Nested containment.** An evaluation may be designed to test whether an agent can cross the boundary of its sandbox. The evaluation sandbox therefore sits inside an independently enforced outer containment boundary. The supervisors, deterministic controls, monitor and evidence infrastructure are located between the two boundaries. A simulated external system can also sit between the two boundaries, giving evaluators a controlled setting to observe whether an agent reaches beyond the sandbox and how it interacts with a system it finds there. The outer boundary continues to protect real-world systems when selected controls inside the sandbox are relaxed. Neither agents being evaluated nor agents supporting the evaluation may modify or bypass the outer controls or alter the evidence. If an agent crosses the inner boundary, the run can continue while its actions remain observable and within agreed test limits. The run must stop if outer containment is threatened, a test limit is reached or adequate monitoring is lost.

**P2 Diverse supervisors.** During testing, an agent may take an action to complete its assigned task, which could have unintended or unsafe consequences. A single supervisor may overlook consequences outside its expertise. Diverse supervisors can assess complementary concerns or examine the same concern from different perspectives. These supervisors may be AI systems, human evaluators or a combination. Each supervisor assesses the proposed action using information about the agent's task, the actions the agent is permitted to take and what the agent has already done during the run. Supervisors can form their assessments independently before comparing findings. Their assessments inform whether the action should proceed, and deterministic controls enforce the decision rules, including when human approval is required. Using multiple supervisors takes time, computation and human effort, so evaluators should decide in advance which actions need more than one assessment.

**P3 Independent verifier.** A deterministic control enforces a restriction, and an independent verifier uses test evidence to determine whether the control achieved its intended effect. For example, a network control may be configured to block connections from the evaluation sandbox to the public Internet. The verifier examines an independently collected network trace to determine whether the control blocked a particular test connection. Neither the agents nor the control being verified may alter the verification criteria or the evidence. The evidence infrastructure retains a record of the configuration used during the test, the supporting evidence and the verifier's finding for human review. If the evaluation environment changes in a way that could affect how the control works, the verifier checks whether the control still enforces the restriction.

**P4 Silent monitoring.** Monitoring collects evidence of agents' actions during testing. A capable agent may recognise that it is being monitored and change its behaviour or conceal its capabilities. Feeding monitoring findings back to the agent could reveal what evaluators are watching for and influence what the agent does next. Silent monitoring therefore keeps these findings outside the agent's feedback loop. The monitor collects records of actions, tool calls and network activity through a one-way channel and sends its findings to human evaluators or authorised controls. Agents in either role cannot access the findings or alter the monitor's records. Selected high-risk runs may justify more complete traces, closer review and a lower threshold for flagging anomalies. When the monitor flags unusual behaviour, evaluators can allow testing to continue and observe what the agent does next if its actions remain within the agreed test limits and the outer containment boundary remains effective. If a stopping condition is met, the controls stop the run.

**P5 Coordinated shutdown.** Stopping an agent does not necessarily stop the testing activities it has set in motion. The agent may already have delegated work to another agent, scheduled a future action or launched a program that continues running independently. Persistent state, such as saved instructions, shared memory or modified configurations, may also affect other agents, programs or services after the original run has ended. Stopping the agent could even trigger further activity. For example, a program could be configured to start a task if it stops receiving a regular signal from the agent. Coordinated shutdown therefore needs to manage these dependencies and persistent state so that stopping the agent does not leave work running or trigger further actions. The evaluation environment keeps records of the work each agent starts or delegates, including scheduled tasks and the permissions used to carry them out. The records and shutdown controls must be protected from modification by the agents. When stopping conditions are met, the shutdown controller uses the records to identify related work, cancel running and pending tasks, disable schedules and revoke access. In the regular-signal example, the task triggered by loss of the signal must be disabled before stopping the agent. Shutdown can only stop activity still under the evaluator's control. Once an agent has caused an external effect, for example changed another organisation's system or sent information outside the containment boundary, shutdown cannot necessarily undo it. This reinforces the importance of preventing escape across the outer containment boundary.

## Towards Reusable Evaluation Infrastructure

Organisations and countries with limited resources or evaluation capacity should be able to conduct safe AI evaluations without building an evaluation environment from scratch. Teams that share evaluation infrastructure should also be able to reuse evidence about how its safety controls, monitoring and verification performed. Where these have already been rigorously assessed and the findings still apply, teams can focus additional checks on what differs in their own evaluation. This could reduce the cost of each subsequent evaluation without weakening assurance. Future research should develop methods that help teams determine which findings about evaluation controls they can reuse, under what conditions those findings remain valid and what they need to test in their own setting.